\documentclass[12pt, preprint]{aastex}
\usepackage{epsfig}    
\usepackage[figuresright]{rotating}  
\usepackage{amsmath}   
\usepackage{amssymb}   
\usepackage{latexsym}  
\usepackage{dcolumn}
\newcolumntype{.}{D{.}{.}{-1}}

\def\vec#1{{\bf #1}}

\begin{document}

\shorttitle{The Hanle Effect in Ly${\alpha}$}

\title{The Hanle Effect of the Hydrogen Ly${\alpha}$ Line for Probing \\the Magnetism of the Solar Transition Region\footnote{\bf To appear in The Astrophysical Journal Letters (2011).}}

\author{Javier Trujillo Bueno\altaffilmark{1,2,3}, Ji\v{r}\'i \v{S}t\v{e}p\'an\altaffilmark{1,2,\dag}, Roberto Casini\altaffilmark{4}}
\altaffiltext{1}{Instituto de Astrof\'{\i}sica de Canarias, 38205 La Laguna,
Tenerife, Spain}
\altaffiltext{2}{Departamento de Astrof\'\i sica, Facultad de F\'{\i}sica, Universidad de La Laguna, Tenerife, Spain}
\altaffiltext{3}{Consejo Superior de Investigaciones Cient\'{\i}ficas (Spain)}
\altaffiltext{4}{High Altitude Observatory, National Center for Atmospheric Research$^{*}$, P. O. Box 3000, Boulder, CO 80307, USA}
\altaffiltext{\dag} {Associate Scientist at Astronomical Institute ASCR, Ond\v{r}ejov, Czech Republic}
\altaffiltext{*} {The National Center for Atmospheric Research is sponsored by the National Science Foundation}
\email{jtb@iac.es, stepan@iac.es, casini@ucar.edu}

\begin{abstract}

We present some theoretical predictions concerning the amplitude and magnetic sensitivity
of the linear polarization signals produced by scattering processes in the hydrogen Ly${\alpha}$ line of the solar transition region. To this end, we have calculated the atomic level polarization (population imbalances and quantum coherences) induced by anisotropic radiation pumping in semi-empirical and hydrodynamical models of the solar atmosphere, taking into account radiative transfer and the Hanle effect caused by the presence of organized and random magnetic fields. The line-center amplitudes of the emergent linear polarization signals are found to vary typically between 0.1\% and 1\%, depending on the scattering geometry and the strength and orientation of the magnetic field. The results shown here encourage the development of UV polarimeters for sounding rockets and space telescopes with the aim of opening up a diagnostic window for magnetic field measurements in the upper chromosphere and transition region of the Sun.

\end{abstract}

\keywords{polarization --- radiative transfer  --- scattering --- Sun: chromosphere --- Sun: surface magnetism}

\section{Introduction}

One of the big challenges of 21st century astrophysics is to understand the magnetism of the Sun, and in so doing to develop the tools needed for the exploration of the magnetic activity in other types of stars across the Hertzsprung-Russel diagram. A crucial step toward this goal is to decipher the magnetic structure of the upper chromosphere, the key interface region between the underlying (cooler) photosphere and the overlying (hotter) corona where dominance of the physics passes from hydrodynamic to magnetic forces (e.g., the review by Harvey 2006). To this end, we need: 

\begin{itemize}

\item to identify observables sensitive to the magnetic fields of the upper chromosphere and transition region; 

\item to develop suitable diagnostic tools to infer the magnetic field from those observables;

\item to design and build the telescopes and instrumentation needed for measuring the observables.

\end{itemize}

The upper chromosphere is indeed a very important part of the Sun because all of the plasma, magnetic field and energy in the corona and solar wind are supplied through this key boundary region. The observables that contain information on the physical conditions of this region are mainly spectral lines in the ultraviolet (UV) and far-UV (FUV) spectral range, with Ly${\alpha}$ being the most prominent line. The intensity profile (i.e., the Stokes $I(\lambda)$ profile) of Ly${\alpha}$ and other lines of the Lyman series have been measured on the solar disk by several instruments on board rockets and space-based telescopes (e.g., Roussel-Dupr\'e 1982; Warren et al. 1998). These measurements show that when observed on the solar disk the hydrogen Lyman lines are always in emission, and that the emission originates in the upper chromosphere and transition region. Interestingly, high-resolution Ly${\alpha}$ {\em intensity} images of quiet-Sun regions taken with the Very high Angular resolution ULtraviolet Telescope (VAULT) sounding rocket show a forest of elongated fibrils suggesting closed magnetic loops in the upper chromosphere (Vourlidas et al. 2010). However, no quantitative information on the magnetic fied of the upper solar atmosphere can be obtained from this type of observations because the intensity profile of Ly${\alpha}$ and of many other transition region lines is practically insensitive to the strength, inclination and azimuth of the magnetic field vector. 

The only way to obtain quantitative empirical information on the strength and orientation of the magnetic field is through the measurement and interpretation of the polarization that some physical mechanisms introduce in spectral lines. Unfortunately, the familiar Zeeman effect is of little practical interest for the ``measurement" of magnetic fields in the upper chromosphere and transition region (except perhaps in sunspots) because the circular polarization (Stokes $V$) scales with the ratio, ${\cal R}$, between the Zeeman splitting and the Doppler line-width and the linear polarization (Stokes $Q$ and $U$) scales with ${\cal R}^2$; 
this ratio is very small for the UV and FUV lines that originate in the weakly magnetized plasma of the upper solar chromosphere because ${\cal R}\,{\propto}\,{\lambda}{B}/{\sqrt{T}}$ (with $\lambda$ the wavelength, $B$ the magnetic strength and $T$ the kinetic temperature; see Landi Degl'Innocenti \& Landolfi 2004). Fortunately, there is yet another physical mechanism by means of which the magnetic fields of a stellar atmosphere leave fingerprints in the spectral line polarization: the Hanle effect. The absorption of {\em anisotropic} radiation produces atomic level alignment (i.e., the individual magnetic substates of levels with angular momentum $j{\ge}1$ are unevenly populated, in such a way that the populations of sublevels with different values of $|M|$ are different and, moreover, quantum coherences between them may appear). This, in turn, gives rise to linear polarization in the spectral line under consideration (e.g., Trujillo Bueno 2001). The Hanle effect due to the presence of a magnetic field inclined with respect to the symmetry axis of the pumping radiation field modifies the atomic level alignment and the linear polarization of the spectral line radiation. Approximately, the emergent linear polarization is sensitive to magnetic strengths between 0.2$B_H$ and 5$B_H$, where $B_H{=}\,{1.137{\times}10^{-7}}/{t_{\rm life}g}$ is the critical Hanle field (in G) for which the Zeeman splitting of the line's level under consideration is similar to its natural width (with $t_{\rm life}$ the level's radiative lifetime, in seconds, and $g$ its Land\'e factor). Note that $B_H{\approx}50$ G for Ly${\alpha}$, $B_H{\approx}20$ G for Ly${\beta}$ and $B_H{\approx}8$ G for Ly${\gamma}$, and that the magnetic sensitivity provided by the Hanle effect is independent of the Doppler line width.

Each Lyman line results from two blended transitions between a lower level, with $j=1/2$, and two upper levels, with $j=1/2$ and $j=3/2$. Therefore, for each Lyman line the only level that can be aligned and contribute to the emergent linear polarization is the upper level with $j=3/2$ (when the hyperfine structure of hydrogen is neglected, which is known to be a good approximation for Ly${\alpha}$). A particularly important question is whether the atomic alignment of the $2p^2{\rm P}_{3/2}$ level of Ly${\alpha}$ is significant or, in other words, if the Ly${\alpha}$ radiation that illuminates the hydrogen atoms of the solar atmosphere is sufficiently anisotropic in the line formation region. Obviously, the anisotropy of the incident Ly${\alpha}$ radiation is very significant in the optically thin case of the solar corona observed off-the-limb at large distances above it (e.g., Trujillo Bueno et al. 2005; and more references therein). The key question for us here is however whether the Ly${\alpha}$ radiation is sufficiently anisotropic within the solar transition region itself, so as to produce there a significant amount of atomic alignment in the $2p^2{\rm P}_{3/2}$ level. As mentioned above, the Lyman radiation that we observe when pointing to the solar disk is produced by the plasma of the solar transition region, which is very optically thick at the central wavelength of Ly${\alpha}$. As a matter of fact, the center-to-limb variation (CLV) of the observed Ly${\alpha}$ intensity is very small or negligible (e.g., Roussel-Dupr\'e 1982; Curdt et al. 2008), which implies that the intensity of the {\em outgoing} radiation (i.e., that propagating outwards) shows no significant variation with the heliocentric angle $\theta$. However, this does not mean that the illumination of the transition region atoms is isotropic -- which would imply that the scattering polarization in Ly${\alpha}$ is zero. As we shall see below, the intensity of the {\em incoming} radiation (i.e., that propagating inwards)
calculated in semi-empirical and hydrodynamical models of the solar chromosphere shows significant CLV, so that the hydrogen atoms of the solar transition region must be illuminated by an anisotropic radiation field. The main goal of this Letter is to show that the 
linear-polarization amplitude of the emergent Ly${\alpha}$ radiation is predicted to be significant because of the anisotropy of the incoming radiation and that via the Hanle effect it is sensitive to the magnetic fields expected for the upper chromosphere and transition region of the Sun.

We focus here on Ly${\alpha}$ because it is the most intense line of the solar transition region and also because its polarization is practically insensitive to collisional depolarization there, but we point out that linear polarization due to atomic level alignment is also expected for at least Ly${\beta}$ and Ly${\gamma}$. 

\section{Formulation of the Problem}

We have applied the quantum theory of spectral line formation described in Chapter 7 of 
Landi Degl'Innocenti \& Landolfi (2004), which treats the scattering line polarization phenomenon as the temporal successsion of 1st-order absorption and re-emission processes, interpreted as statistically independent events. This complete frequency redistribution (CRD) approximation is suitable for reaching the objective of this Letter -- that is, to estimate the {\em line-center} polarization amplitudes in the absence and in the presence of the Hanle effect (e.g., Sampoorna et al. 2010, and more references therein). 

The splitting between the $2p^2{\rm P}_{1/2}$ and $2p^2{\rm P}_{3/2}$ upper levels of Ly${\alpha}$ 
is of the order of $10^{10}\,{\rm s}^{-1}$, which is much larger than the level's natural width (of the order of $10^8\,{\rm s}^{-1}$) but much smaller than the line's Doppler width (of the order of $10^{11}\,{\rm s}^{-1}$). This justifies to neglect quantum interferences between such two $j$-levels (see Belluzzi \& Trujillo Bueno 2011). 
The upper level $2p^2{\rm P}_{3/2}$, with Land\'e factor $g=4/3$, is the only level that contributes to the linear polarization of Ly${\alpha}$ when its hyperfine splitting is neglected, which as shown by Bommier \& Sahal-Br\'echot (1982) is a good approximation.

In order to estimate the linear polarization amplitudes of Ly${\alpha}$ we need to take into account radiative transfer effects in models of the solar atmosphere. To this end, we have chosen both the quiet-Sun semi-empirical model of Fontenla et al. (1993; hereafter the FAL-C model) and the chromospheric hydrodynamical model of Carlsson \& Stein (1997; hereafter the hydrodynamical model). The atomic model we have used includes all the fine-structure levels of the first three $n$-levels of hydrogen (that is, it includes Ly${\beta}$ and H${\alpha}$, in addition to Ly${\alpha}$). We quantify the excitation state of each $j$-level by means of the multipolar components of the atomic density matrix, whose self-consistent values at each spatial grid point of the model atmosphere have to be obtained by solving jointly the statistical equilibrium equations for such density-matrix components and the Stokes-vector transfer equation for each of the allowed radiative transitions in the atomic model. To this end, we have applied an efficient and accurate  radiative transfer method (see Appendix A of \v{S}t\v{e}p\'an \& Trujillo Bueno 2011). Isotropic collisions with protons and electrons have been also taken into account, as described in Appendix B of \v{S}t\v{e}p\'an \& Trujillo Bueno (2011). In particular, the dipolar collisional rates between nearby $j$-levels (i.e., between levels $nlj$ and $nl\pm1j^{'}$) have been calculated applying the semi-classical impact approximation theory (e.g., Sahal-Br\'echot et al. 1996), while the inelastic collisional rates between the different $n$-levels were taken from Przybilla \& Butler (2004). We have included also the Stark broadening of the hydrogen lines, following Stehl\'e (1996). With these physical ingredients, our non-LTE syntheses of the hydrogen intensity profiles are in good agreement with those computed by other researchers using more $n$-levels.

\section{The anisotropy of the Ly${\alpha}$ radiation in the solar transition region}

In the absence of magnetic fields the line-center amplitude of the fractional scattering polarization in the Lyman lines 
can be estimated by applying the following approximate expression (see Trujillo Bueno \& Manso Sainz 1999, and note that 
the $1/\sqrt{2}$ factor here takes into account that two blended transitions contribute to the intensity profile):

\begin{equation}
\frac QI\,\approx\,{1\over{\sqrt{2}}}(1-\mu^2)\,{\rm W}_2(j_l,j_u)\,{{{\bar J}^2_0}\over{{\bar J}^0_0}}\,,
\label{eq:qieddb}
\end{equation}
where ${\rm W}_2(j_l,j_u)=1/2$ for the Lyman lines, $\mu={\rm cos}\,{\theta}$ (with $\theta$ the heliocentric angle), and ${{\bar J}^2_0}/{{\bar J}^0_0}$ is the fractional anisotropy of the spectral line radiation calculated at the height in the model atmosphere where the line center optical depth is unity along the line of sight (LOS). 
Therefore, the key quantity that determines the scattering polarization signals we may expect for the Ly${\alpha}$ line when observing the solar disk is ${{\bar J}^2_0}/{{\bar J}^0_0}$, 
where 

\begin{equation}
{\bar J}^0_0=\int d{\nu}
\oint \frac{d \vec{\Omega}}{4\pi}\,{{\phi}}({\nu},\vec{\Omega})\,{{I_{{{\nu}} \vec{\Omega}}}}\,
\label{eq:j00}
\end{equation}
is the familar frequency-integrated mean intensity ($\phi({\nu},\vec{\Omega})$ being the normalized absorption profile, with $\nu$ the frequency and $\vec{\Omega}$ the ray direction),  and 

\begin{equation}
{\bar J}^2_0=\frac{1}{2\sqrt{2}} \int d{\nu}
\oint \frac{d \vec{\Omega}}{4\pi}\,{{\phi}}({\nu},\vec{\Omega})\,\left[(3\mu^2-1){{I_{{{\nu}} \vec{\Omega}}}}+
3(\mu^2-1){Q_{{{\nu}} \vec{\Omega}}}\right]\,
\label{eq:j20}
\end{equation}
quantifies the anisotropy of the spectral line radiation that illuminates each point of the astrophysical plasma under consideration\footnote{Note that these are the CRD expressions for ${\bar J}^K_0$ (with $K=0,2$). Partial frequency redistribution expressions for ${\bar J}^K_0$ can be found in the paper by Sampoorna et al. (2010), which clarifies that CRD is a suitable approximation for estimating the line-center scattering polarization.}. In solar-like atmospheres the first term of Eq. (3) plays the dominant role, so ${\bar J}^2_0{>}0$ if the incident radiation field is predominantly vertical while ${\bar J}^2_0{<}0$ if it is predominantly horizontal. The specific intensity depends on the spatial variation of the corresponding source function component, $S_I$. If $S_I$ decreases with height in the solar atmosphere then the outgoing radiation shows limb darkening (i.e., it is predominantly vertical), while the incoming radiation shows limb brightening (i.e., it is predominantly horizontal). It is the magnitude of the vertical gradient of $S_I$ what determines the sign of ${\bar J}^2_0$ (see Fig. 4 of Trujillo Bueno 2001). For a constant source function we have that ${\bar J}^2_0{<}0$, because in this particular case all outgoing rays have the same intensity and we are left only with the (predominantly horizontal) contribution of the incoming radiation. 

The left panel of Fig. 1 shows the spatial variation of ${{\bar J}^2_0}/{{\bar J}^0_0}$ for H${\alpha}$, Ly${\alpha}$ and Ly${\beta}$, calculated in the FAL-C model. Note that the anisotropies of the Lyman lines are practically zero all through the model atmosphere, except in the model's transition region where they are negative and significant (i.e., of the order of a few percent at the atmospheric heights where the line-center optical depth is unity along the LOS). The right panel of Fig. 1 illustrates that in the model's transition region the Ly${\alpha}$ anisotropy is dominated by the CLV of the incoming radiation. 

\section{Scattering polarization and Hanle effect in Ly${\alpha}$}

The two panels of Fig. 2 show the calculated
center-to-limb variation of the emergent $Q(\lambda)/I(\lambda)$ profile of the Ly${\alpha}$ line (hereafter, $Q/I$).  
The left panel corresponds to the results of our radiative transfer calculations  
in the semi-empirical FAL-C model, while the right panel shows 
the $Q/I$ profiles that result from averaging the time-dependent 
Stokes $I$ and $Q$ profiles calculated in the 
hydrodynamical simulation model of Carlsson \& Stein (1997). 
The variation with $\mu$ of the line center amplitude of the 
$Q/I$ profile can be easily understood by using Fig. 1 (right panel) and Eq. (1). 
Remarkably, the amplitude and shape of the $Q/I$ profiles calculated
in both solar atmospheric models are qualitatively similar; this re-inforces our conclusion that
scattering processes in the solar transition region should indeed produce measurable
linear polarization signals in Ly${\alpha}$. We point out that the transition region starts at a rather precise height in the FAL-C model (i.e., at about 2200 km), while in the hydrodynamical model it fluctuates with time.

We turn now to estimating the sensitivity of the emergent Ly${\alpha}$ linear polarization to the presence of organized and random magnetic fields in the solar transition region. Since similar conclusions are obtained with the hydrodynamical model, 
it suffices with showing the results of our self-consistent radiative transfer calculations in the FAL-C model atmosphere.

Figure 3 
shows the calculated $Q/I$ and $U/I$ signals for a LOS 
with $\mu=0.3$, which implies a distance of about 45\arcsec\ 
from the solar limb.  Except for the $B=0$\,G case (shown in 
Fig. 3 with a gray dotted 
line for reference), the curves show $Q/I$ and $U/I$ for a 
horizontal (i.e., parallel to the solar surface) magnetic field 
of 20 G; the field has a fixed orientation given by the 
azimuth angle written in the figure panels. 
Note that both $Q/I$ and $U/I$ are sensitive to the azimuth of the magnetic field 
vector: while the $Q/I$ signal changes but remains always negative, the sign of 
$U/I$ also depends on the azimuth value. 

In such close-to-the-limb scattering geometry, 
$U/I$ would be zero if, instead of having a dominant orientation, the 
magnetic field azimuth was uniformly distributed within the 
spatio-temporal resolution element of the observation.
However, away from the solar disk center, $Q/I$ remains sensitive to the magnetic field strength even in such an unfavorable situation. This can be seen clearly in the left panel of Fig. 4, which shows how the amplitude of $Q/I$ signal 
decreases as the magnetic strength of a random-azimuth horizontal field 
increases. As seen in this figure, for the scattering geometry of a close to the 
limb observation (e.g., $\mu=0.3$), the Hanle effect {\em depolarizes}.  
However, for the forward-scattering geometry of a disk center 
observation (i.e., a LOS with $\mu=1$), the Hanle effect of an inclined magnetic 
field with a given orientation {\em creates} linear polarization (see the 
right panel of Fig.~4 and note that in order to be able to detect the presence of a horizontal magnetic field of 30 G in the bulk of the solar transition region the polarimetric sensitivity of the measurement should be at least 0.1\%). Therefore, the Hanle effect in Ly${\alpha}$ can be used to diagnose the magnetic field all over the solar disk (i.e., it is not restricted to an annular region around the solar limb).

\section{Conclusions}

Our radiative transfer investigation in semi-empirical and hydrodynamical models of the solar atmosphere indicate that the Ly${\alpha}$ line should show measurable linear polarization signals when observing the solar disk. Via the Hanle effect the line-center amplitudes turn out to be sensitive to the presence of magnetic fields in the solar transition region, with good sensitivity to field strengths between 10 and 100 G. For such field strengths there is no significant impact of the Zeeman effect on the Ly${\alpha}$ radiation.

The above-mentioned prediction is based on the same quantum theory that explains the linear polarization profiles observed in various visible and near-IR lines (e.g., Manso Sainz \& Trujillo Bueno 2003; \v{S}t\v{e}p\'an \& Trujillo Bueno 2010). This CRD theory is suitable also for estimating the line-center polarization amplitudes of strong resonance lines like Ly${\alpha}$, but we point out and advance here that partial frequency redistribution (PRD) effects do produce interesting and sizable $Q/I$ signals in the Ly${\alpha}$ wings (qualitatively similar to those shown in figure 9 of Trujillo Bueno 2011 for the Mg {\sc ii} k-line at 2795 \AA).

The predicted line-center amplitudes of the Ly${\alpha}$ scattering polarization are smaller than 1\%. Nevertheless, with a moderate aperture telescope they can be measured with a polarimetric sensitivity of 0.1\% and a spectral resolution of 0.1 \AA\ by opting for a spatial resolution of the order of 1 arcsecond and a temporal resolution of the order of 1 minute (e.g., see the Ly${\alpha}$ polarization sounding rocket project outlined in Ishikawa et al. 2011). Information on the thermal and magnetic structure of the solar transition region can be obtained from the observed $Q/I$ and $U/I$ profiles themselves and through detailed radiative transfer simulations in given atmospheric models. Therefore, it would be worthwhile to develop a spectropolarimeter for measuring from space the linear polarization profiles caused by scattering processes and the Hanle effect in the Ly${\alpha}$ line of the solar transition region.
Unfortunately, the pioneering Soviet satellite experiment of Stenflo et al. (1980) failed because the optical transmission of the instrument was severely degraded due to in-orbit contamination from the satellite. Such a development would open up a novel diagnostic window for the exploration of the magnetism of the outer solar atmosphere.

{\bf Acknowledgments}
We are grateful to S. Tsuneta (NAOJ), K. Kobayashi (UAH) and all the other members of the Chromospheric Ly${\alpha}$ Spectropolarimeter (CLASP) sounding rocket project for their interest in the theoretical results presented here and for stimulating scientific discussions. Thanks are also due to R. Manso Sainz (IAC) and L. Belluzzi (IAC) for their careful review of this letter, and to M. Carlsson (UiO) for kindly providing the hydrodynamical model atmosphere. Financial support by the Spanish Ministry of Science through projects AYA2010-18029 (Solar Magnetism and Astrophysical Spectropolarimetry) and CONSOLIDER INGENIO CSD2009-00038 (Molecular Astrophysics: The Herschel and Alma Era) is gratefully acknowledged.


\newpage


\begin{figure}[t]
  \centering
\includegraphics[width=8.cm]{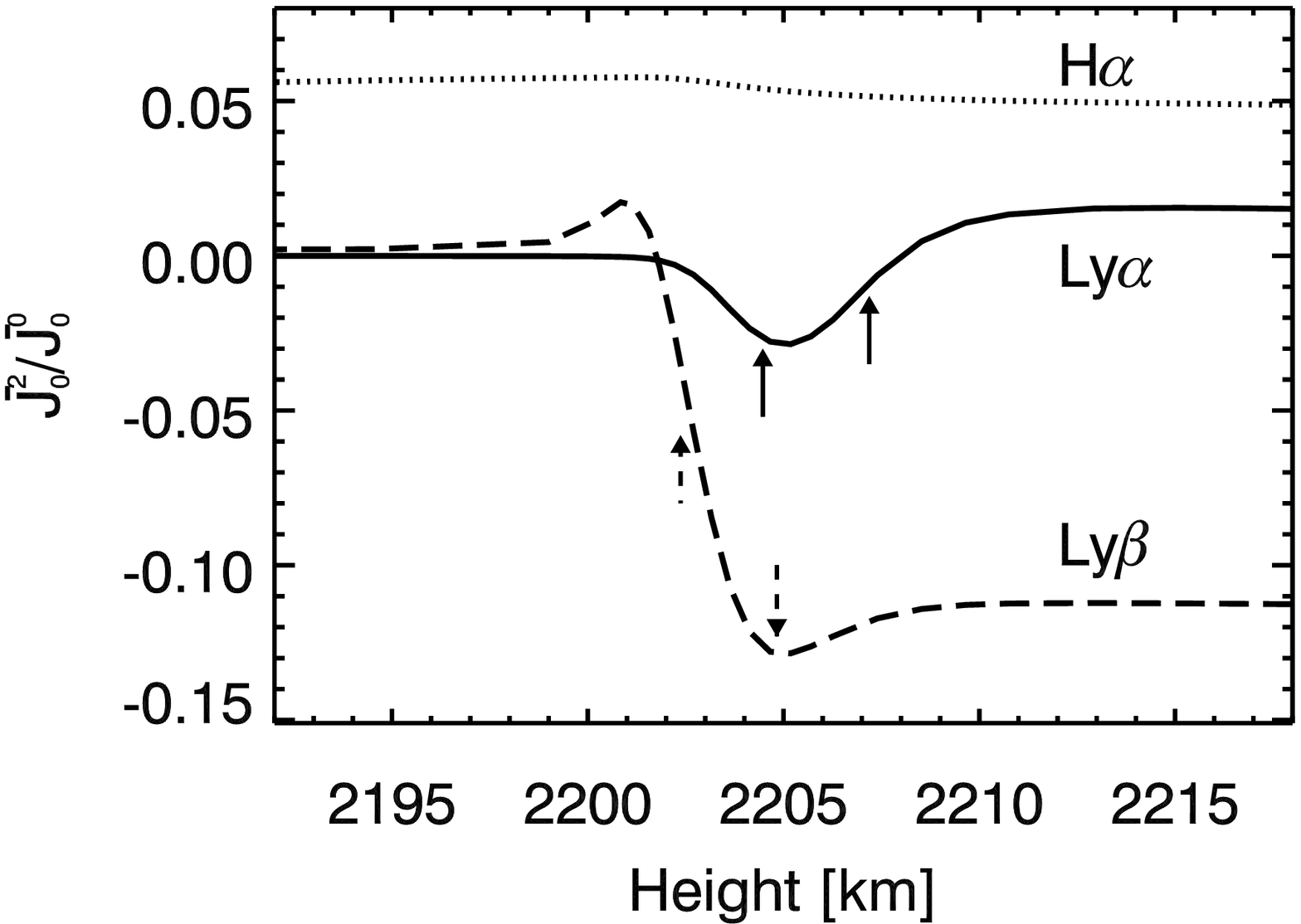}  
\includegraphics[width=8.cm]{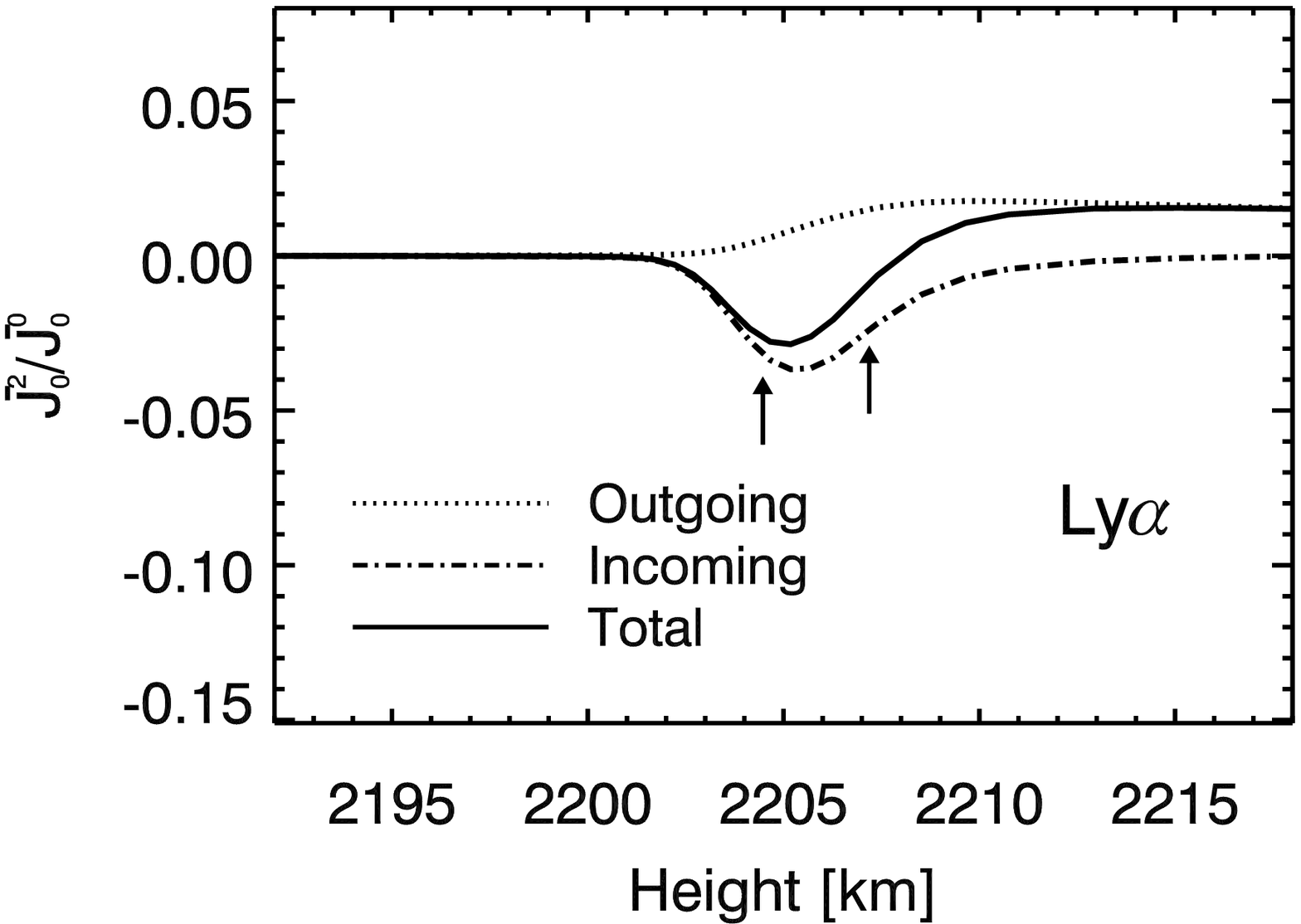}
  \caption[]{The fractional anisotropy of the spectral line radiation. Left panel: the variation with height in the FAL-C model atmosphere of the CRD fractional anisotropy of the Ly${\alpha}$ (solid line), Ly${\beta}$ (dashed line) and H${\alpha}$ (dotted line) radiation, with the arrows indicating the heights where the line center optical depth of each Lyman line is unity along line of sights with $\mu=1$ (left arrow) and $\mu=0.1$ (right arrow). Right panel: like in the left panel, but only for Ly${\alpha}$ and showing also the contributions of the outgoing radiation (with $0{<}{\mu}{\le}1$) and incoming radiation (with $-1{\le}{\mu}{<}0$).}
  
\label{fig:figure-1}
\end{figure}

\clearpage

\begin{figure}[t]
  \centering
\includegraphics[width=8.cm]{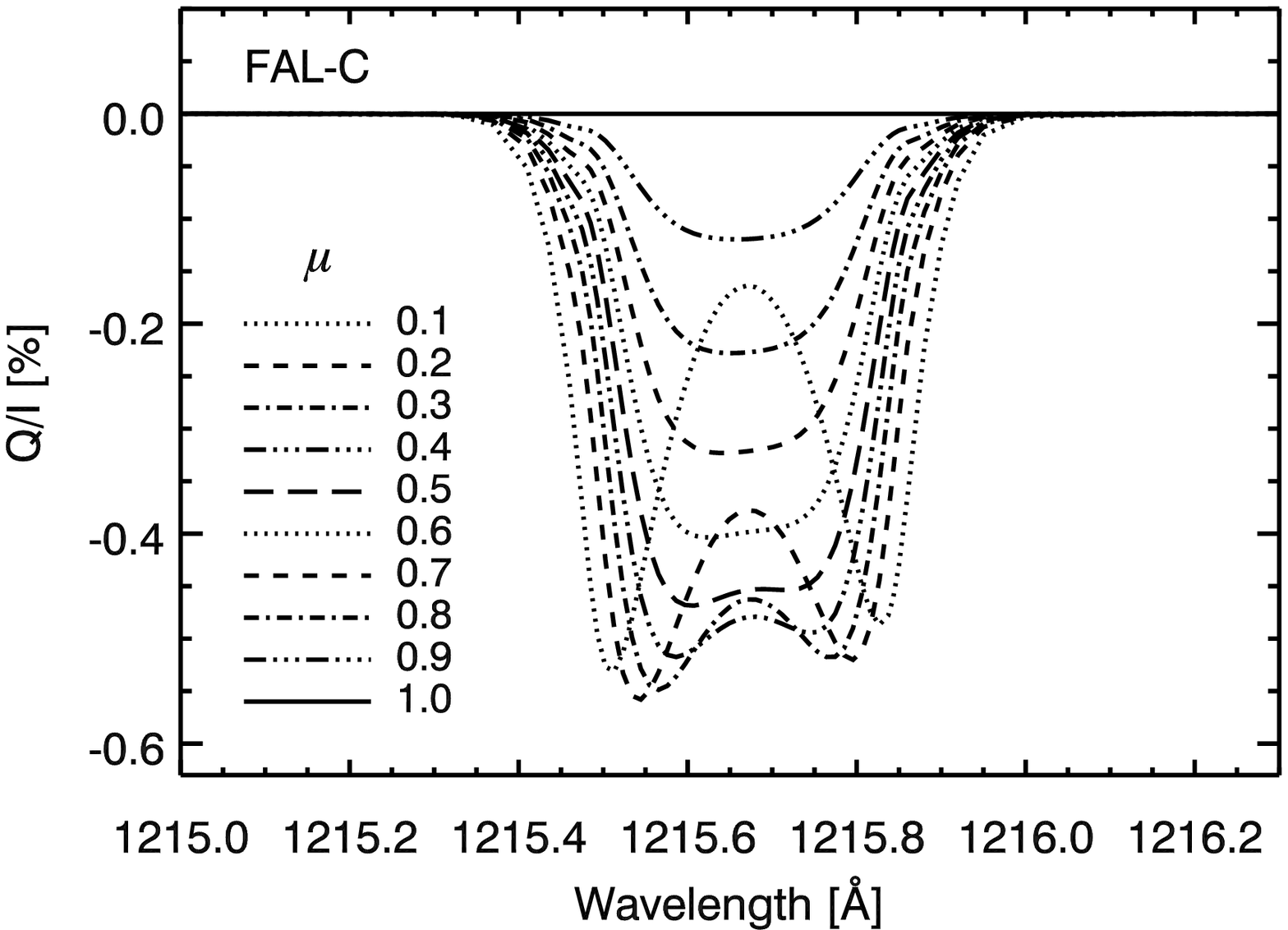}
\includegraphics[width=8.cm]{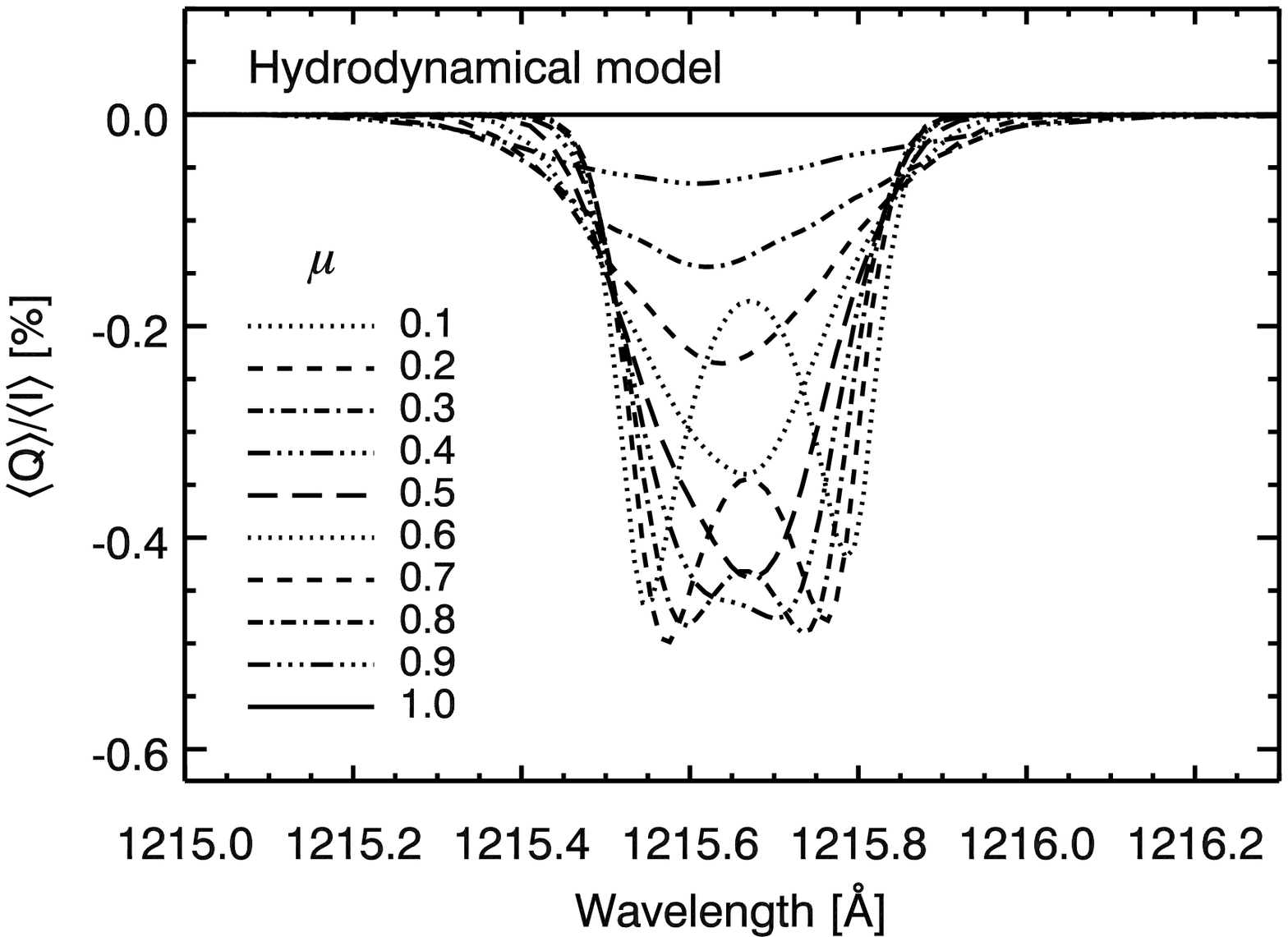}
  \caption[]{The unmagnetized case. Left panel: CLV of the emergent Ly${\alpha}$ $Q/I$ profile calculated in the FAL-C model. Right panel: CLV of the emergent Ly${\alpha}$ $Q/I$ profile that results after time averaging the Stokes $I$ and $Q$ profiles computed at each time step of the hydrodynamical model. The positive reference direction for Stokes $Q$ is the parallel to the nearest limb.}
\label{fig:figure-2}
\end{figure}

\clearpage 

\begin{figure}[t]
  \centering
\includegraphics[width=8.cm]{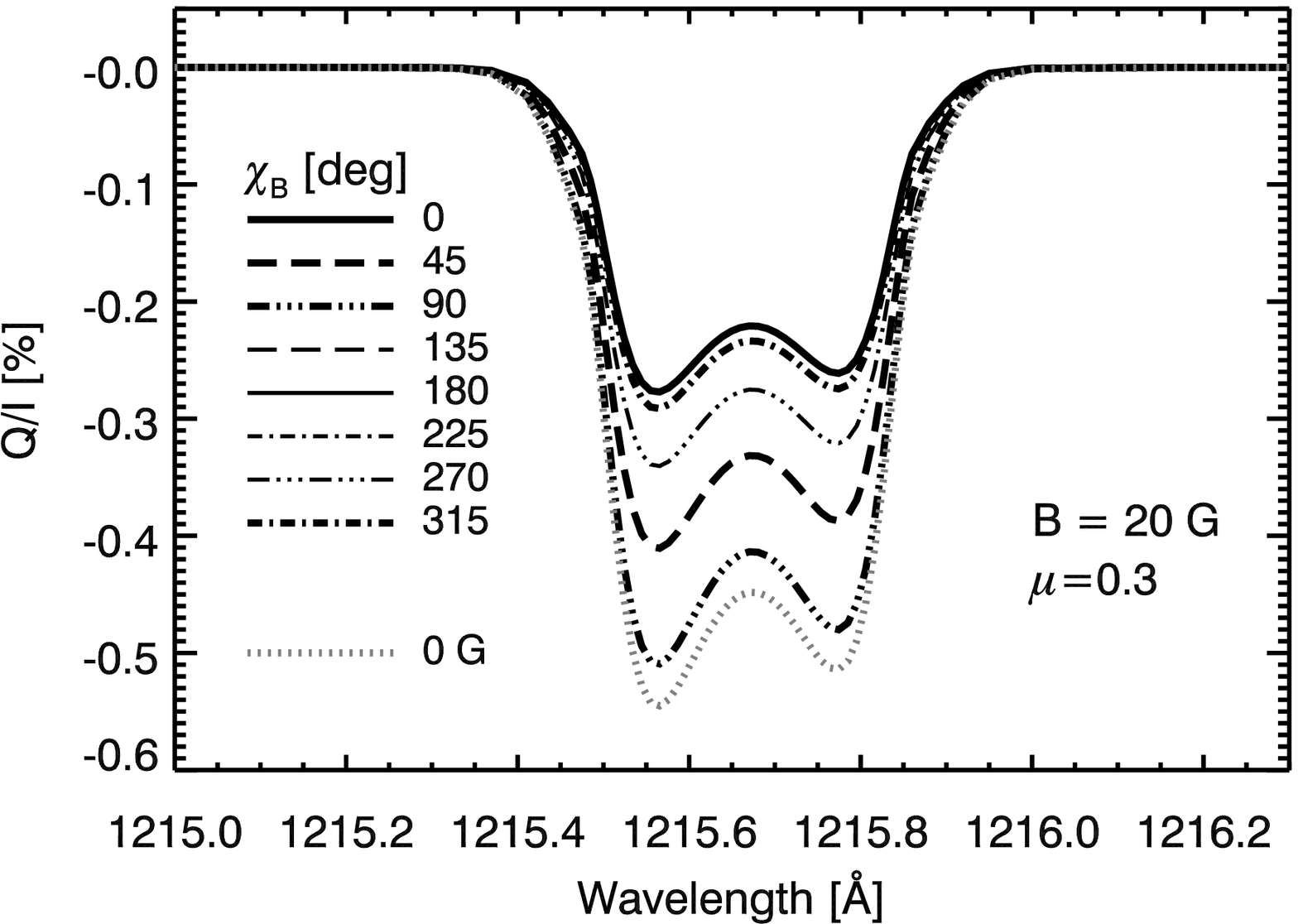}
\includegraphics[width=8.cm]{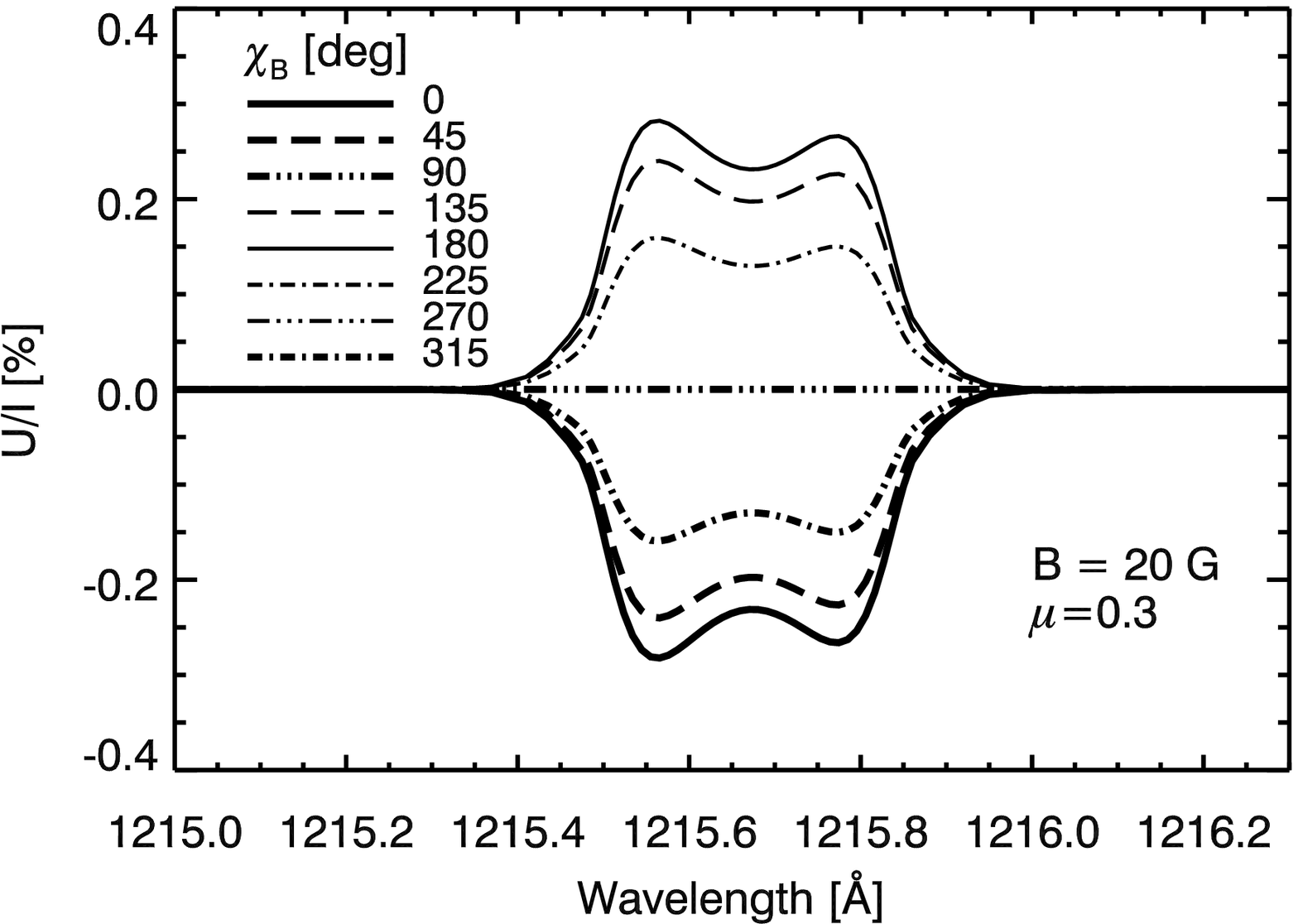}
  \caption[]{The magnetized case. The Ly${\alpha}$ $Q/I$ and $U/I$ signals for a LOS with $\mu=0.3$
produced by scattering processes in the transition 
region of the FAL-C model atmosphere, taking into account the Hanle effect of a 20\,G horizontal magnetic 
field. The various profiles correspond to the indicated values of the magnetic field azimuth ($\chi_B$), 
measured counterclockwise with respect to the projection of the LOS onto the solar surface plane. 
The positive reference direction for Stokes $Q$ is the parallel to the nearest limb.}
\label{fig:figure-3}
\end{figure}

\clearpage

\begin{figure}[t]  
  \centering
\includegraphics[width=8cm]{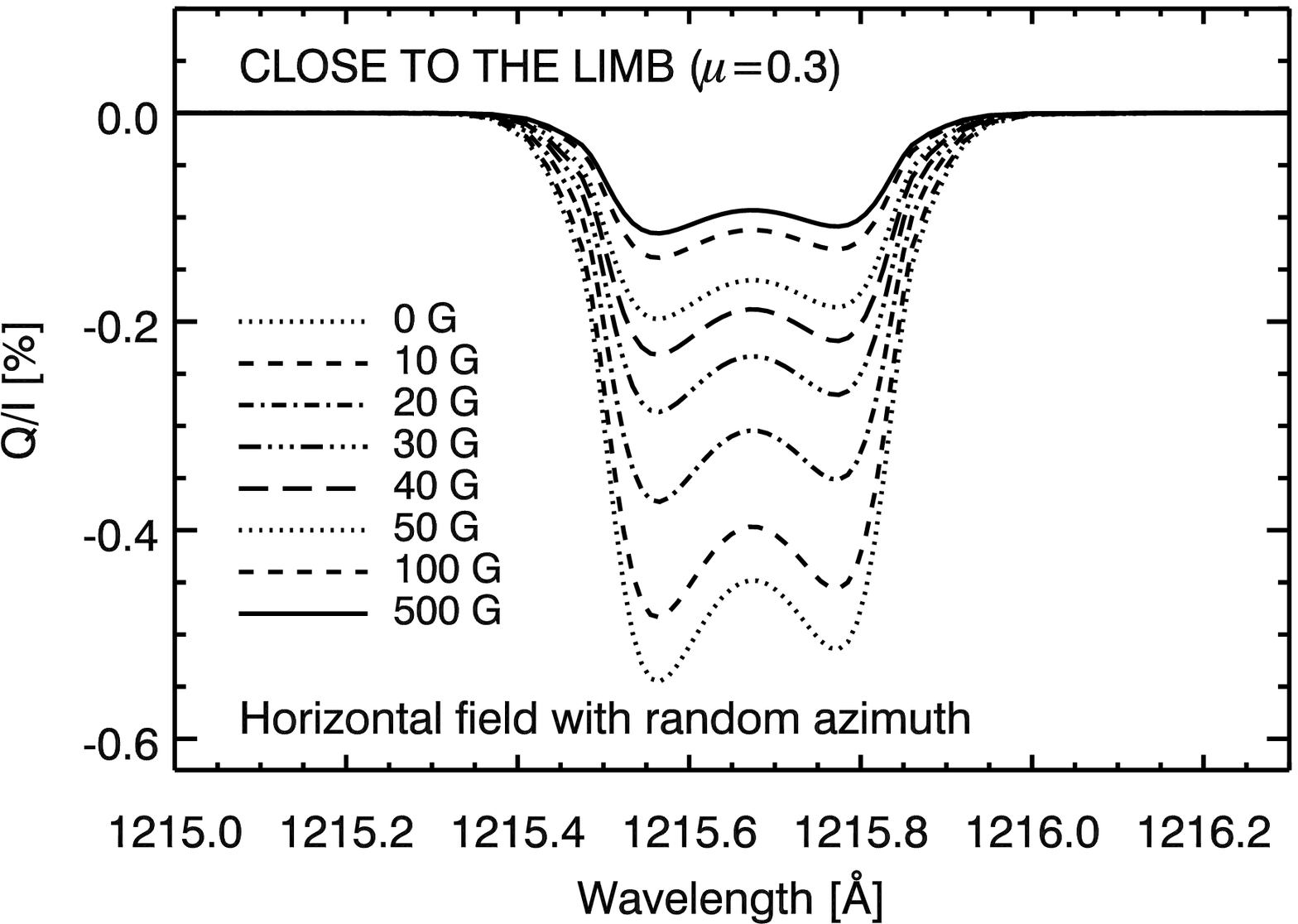}
\includegraphics[width=8cm]{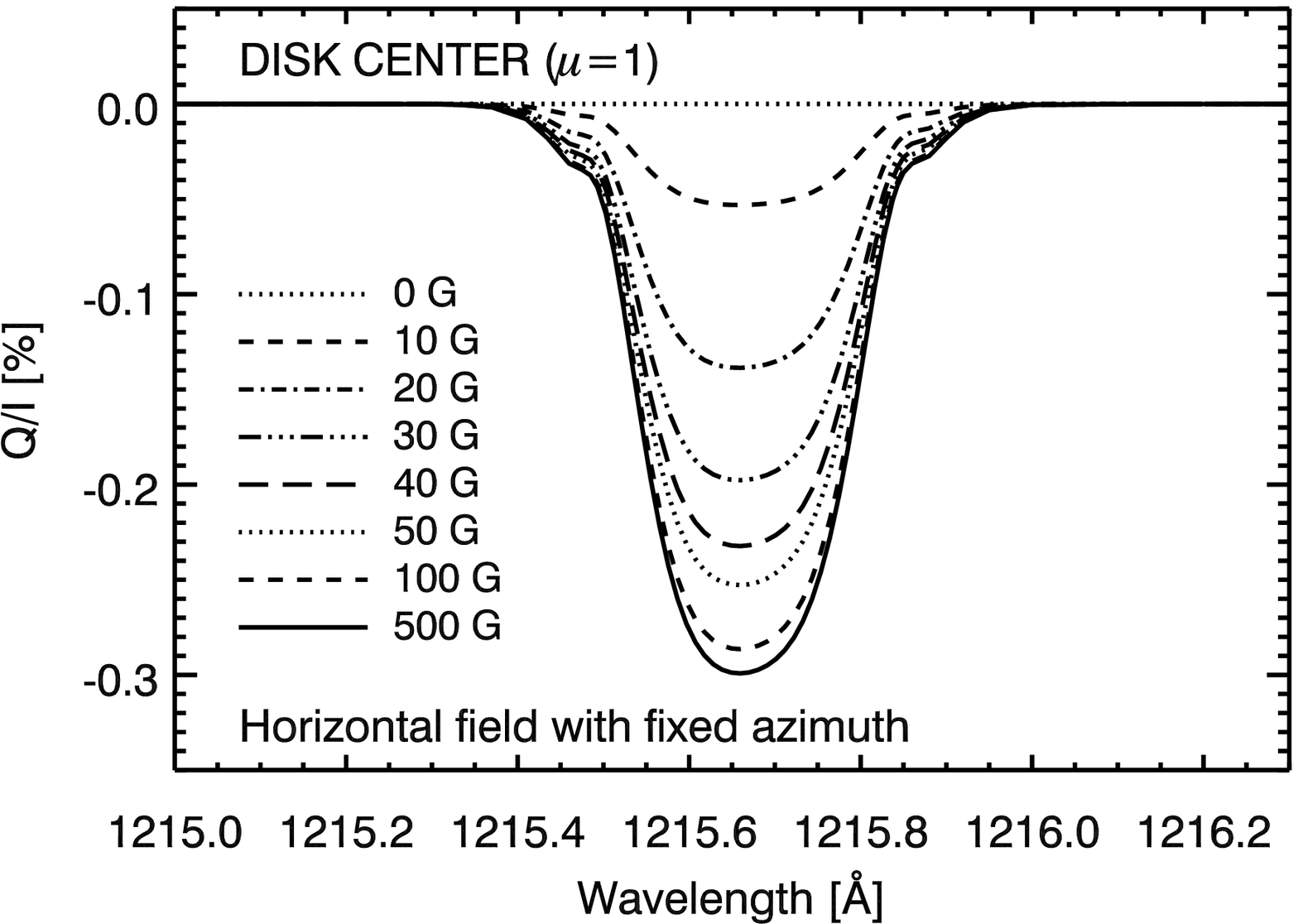}
  \caption[]{The magnetized case. In both panels the various $Q/I$ profiles 
correspond to the indicated values of the magnetic field strength.
  \emph{Left:} The Ly${\alpha}$ scattering 
  polarization signals for a LOS with $\mu=0.3$ assuming the presence in the FAL-C model of a horizontal 
magnetic field with a {\em random} azimuth within the observational resolution 
element. The positive 
reference direction for Stokes $Q$ is the parallel to the nearest limb. 
Note that in this close to the limb scattering geometry the linear polarization amplitude {\em decreases} as the magnetic strength is increased. \emph{Right:} The Ly${\alpha}$ scattering polarization signals for a LOS with $\mu=1$ assuming the presence of a horizontal magnetic field with a {\em fixed} azimuth. The positive 
reference direction for Stokes $Q$ is along the magnetic field.
Note that in this forward-scattering geometry the linear polarization amplitude {\em increases} as the magnetic strength is increased.
}
\label{fig:figure-4}
\end{figure}

\end{document}